\documentclass[prl,showpacs,amsmath,preprint]{revtex4}  
\newcommand{\beq}{\begin{equation}}
\newcommand{\enq}{\end{equation}}

\usepackage{graphicx}

\begin{document}
\title{How much laser power can propagate through fusion plasma?}

\author{Pavel M. Lushnikov$^{1,2,3}$ and Harvey A. Rose$^3$
}

\affiliation{
  $^1$ Landau Institute for Theoretical Physics, Kosygin St. 2,ÄÄ
  Moscow, 119334, Russia\\
$^2$Department of Mathematics, University of  Notre Dame, Indiana
46556, USA \\
$^3$Theoretical Division, Los Alamos National Laboratory,
  MS-B213, Los Alamos, New Mexico, 87545, USA
}

\email{har@lanl.gov}

\date{
\today
}

\begin{abstract}
Propagation of intense laser beams is crucial for inertial
confinement fusion, which requires precise beam control to
achieve the compression and heating necessary to ignite the
fusion reaction. The National Ignition Facility (NIF), where
fusion will be attempted, is now under construction.  Control of
intense beam propagation may be ruined by laser beam
self-focusing.  We have identified the maximum laser beam power
that can propagate through fusion plasma without significant
self-focusing  and have found excellent agreement with recent
experimental data, and suggest a way to increase that maximum by
appropriate choice of plasma composition with implication for NIF
designs.  Our theory also leads to the prediction of
anti-correlation between beam spray and backscatter and suggests
the indirect control of backscatter through manipulation of
plasma ionization state or acoustic damping.
\end{abstract}


\maketitle

\section{Introduction}

Propagation of intense laser beams in plasma raises outstanding
technological and scientific issues.  These issues are closely
tied with inertial confinement fusion (ICF)
\cite{McCrory1988,Still2000,Miller2004,Lindl2004} which requires
precise beam control in order to maintain symmetry of spherical
target implosion, and so achieve the compression and heating
necessary to ignite the fusion reaction. ICF will be attempted at
the National Ignition Facility (NIF). While most engineering
features of NIF are now fixed, there are still crucial choices to
be made \cite{Lindl2004} in target designs. Control of intense
beam propagation is endangered by laser beam self-focusing, when
a beam digs a cavity in plasma, trapping itself, leading to
higher beam intensity, a deeper cavity, and so on.

Self-focusing occurs when an intense laser beam propagates
through a wide range of optical media \cite{Boyd2002}, and has
been the subject of research for more than forty years, since the
advent of lasers \cite{Sulem1999}.    In laser fusion the
intensity of laser beams is so large that self-focusing in plasma
can cause disintegration of a laser beam into many small beams,
leading to rapid change in beam angular divergence $\triangle
\theta$, called beam spray. Significant beam spray is absolutely
unacceptable for attaining fusion which requires precise laser
beam control \cite{Lindl2004}.   It was commonly assumed that the
main source of beam spray in fusion plasma is the self-focusing in
local maxima of laser intensity (hot spots) which are randomly
distributed throughout the plasma \cite{Lindl2004}. Hot spot
self-focusing can be controlled by  reducing beam correlation
time,  $T_c$. However we show in this Article that  the main
limitation of  maximum beam power, which can propagate in plasma
without significant beam spray, is determined by  collective
instability which couples the beam to an ion acoustic wave. We
call this instability {\it collective} forward stimulated
Brillouin scatter (CFSBS) \cite{LushnikovRosePRL2004} because it
does not depend on the dynamics of isolated hot spots, but rather
the intensity fluctuations as temporally smoothed (averaged) by
ion inertia. We show below that this collective instability is
consistent with the first experimental observation of the beam
spray onset \cite{NiemannPRL2005} while hot spot self-focusing is
not.

\section{beam collapse (catastrophic self-focusing)}

There are two self-focusing mechanisms in plasma: ponderomotive
and thermal. Historically, ponderomotive self-focusing was
studied first.   The ponderomotive mechanism results from
averaging over fast electron oscillations in the laser
electromagnetic field, at frequency $\omega_0$. Averaging induces
an effective electrostatic potential proportional to the local
laser intensity, which in turn adds to the usual fluid pressure
term in hydrodynamical equations \cite{Kruer1990}. The thermal
mechanism results from the transport of electron temperature
fluctuations, $\delta T_e.$

Ponderomotive self-focusing in three dimensions (3D) is quite
different than in two dimensions (2D). (Here one dimension is the
direction of propagation of laser beam with one/two transverse
dimensions in 2D/3D, respectively).  In 2D, self-focusing often
results in propagation of optical pulses (called solitons
\cite{ZakharovShabat1971}) without change of their shape over
large distances. In 3D, self-focusing often leads to dramatic
intensity amplification with propagation distance. Indeed,
self-focusing of light, as described by the nonlinear Schrodinger
equation, results in formation of a point singularity after
finite distance of light propagation
\cite{ChiaoGarmireTownes1964,TalanovLETPLett1965}. A finite
amount of optical power is drawn into this point, usually
referred to as beam collapse. Near singularity, the nonlinear
Schrodinger equation looses its applicability because of finite
density depletion effects and instead of singularity, light
scatters in a wide range of angles, causing loss of precise
irradiation symmetry necessary for fusion.  For application to
fusion, only the 3D regime is relevant, and only this regime is
considered in this Article. Note that in some regimes other, high
frequency instabilities, such as stimulated Raman scatter can
also arrest catastrophic collapse (see e.g. Ref.
\cite{RoseDuBoisPRL1994}) but they are not considered here.

Beam collapse occurs if the laser beam power, $P$, exceeds a
critical value \cite{TalanovLETPLett1965},  $P_c\propto T_e/n_e$.
$T_e$ and $n_e$ are the electron temperature and density,
respectively. For NIF parameters ($n_e\approx 10^{21}/cm^3, \
T_e\approx 5 \mbox{keV}, \ \omega_0\approx 5\times
10^{15}\mbox{sec}^{-1})$ $P_c=1.6 \times 10^9$ Watts. This power
evaluation is based on Ref. \cite{TalanovLETPLett1965}, in
contrast to threshold given by Max \cite{Max1976}, which is
roughly half as large.  The former may be dynamically realized
(see Eq. (107) of Ref. \cite{RoseDuBoisPhysFluid1993}) from
non-equilibrium initial conditions, appropriate to initiation by
hot spots, while the latter is strictly an equilibrium property,
and hence not useful for quantitative beam propagation prediction.

The energy required for inertial confinement fusion is so large
that the power in each of  NIF's 48 beam quads \cite{Lindl2004}
exceeds $P_c$  by several orders of magnitude: the power of each
NIF beam is approximately $8 \times 10^{12}$ Watts, or about $5
\times  10^3$ critical power. This difficulty is alleviated by the
Random Phase Plate (RPP) \cite{KatoMima1982} which splits the
laser beam into many (tens of thousands) small beams with random
phases, which are then optically focused into plasma (see Figure
2 in Ref. \cite{Still2000}). As a result the total laser beam
electric field amplitude, $E$, is well approximated in vacuum as
an anisotropic random Gaussian field, with correlation length
$l_c$ perpendicular to the beam propagation direction, much
smaller than the parallel correlation length. The laser
intensity, $I\propto |E|^2$, forms a speckle field - a random in
space distribution of intensity (see Figure 1a).
\begin{figure}
\begin{center}
\includegraphics[width = 4.4 in]{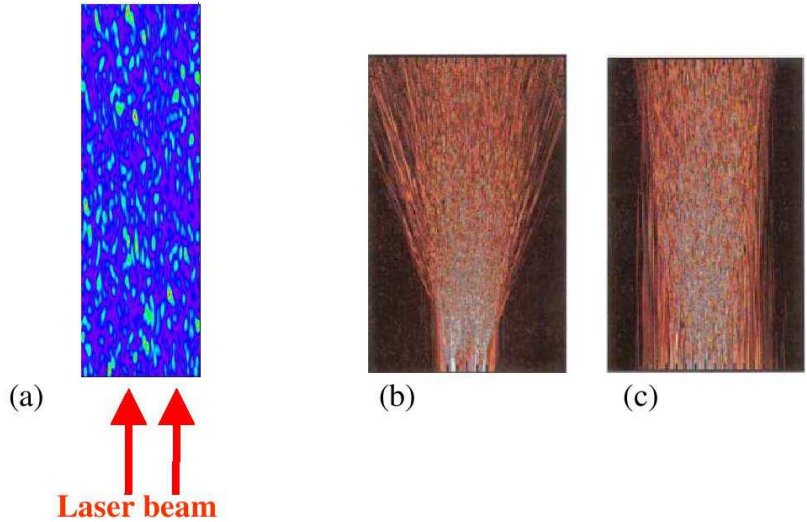}
\caption{Figure 1. Two dimensional slice of light intensity
fluctuations inside plasma. Laser beam propagates from bottom of
figure upward. (a) Distribution of fluctuations at scale much
smaller than the beam diameter. Random  fluctuations
(``speckles"), are highly anisotropic, with correlation or
speckle length along the beam propagation, $``z"$, direction
about $7F^2\lambda_0$. (b) Beam spray regime of laser
propagation. Beam disintegrates into many small beams. (c)
Negligible beam spray regime. This regime is necessary for
attaining fusion (from Ref. \cite{Still2000} with permission).
Horizontal scale in (b) and (c) correspond to beam diameter.}
\label{fig:fig1}
\end{center}
\end{figure}

\section{Time-independent self-focusing}

First consider the regime where laser beam time dependence is
negligible. If the average intensity, $\langle I\rangle $, is
small, then collapse events occur only in speckles (also referred
to as hot spots) with  $I\gg \langle I\rangle$, so that their
power,  $P\sim l_c^2I$, exceeds $P_c$. The width of these intense
speckles, $F\lambda_0$, is much smaller than the beam diameter and
is determined by the laser optic system, where $\lambda_0$ is the
laser wavelength in vacuum and $F$ is the optic $f-$number (the
ratio of the focal length of the lens divided by the lens
diameter). We take $l_c=F\lambda_0/\pi$. Since there is always
finite probability of obtaining such collapsing speckles in the
random Gaussian field model, the beam angular divergence,
$\triangle \theta$, increases with each collapse event.
$\triangle \theta$ in vacuum is given by $\triangle \theta=1/F$,
for $F^2\gg 1$. If the probability of speckle collapse is small,
then the beam will keep its initial form.  But if laser power if
so large that power of many hot spots exceeds $P_c$ then the beam
will disintegrate into many small beams, leading to rapid change
in  $\triangle \theta$, (beam spray). Figures 1b and 1c show
examples of both regime of strong and negligible beam spray.

An important measure of beam spray in this time independent
regime is the fraction,  $P_{scattered}$, of beam power,
$P_{beam}$, in speckles which self-focus as the beam propagates,
estimated as follows. NIF optic is approximately square, and
hence a speckle area is  $F^2\lambda_0^2$, implying a critical
intensity for speckle self-focusing,
$I_c=P_c/(F\lambda_0)^2\approx 2\times 10^{16}
\mbox{W}/\mbox{cm}^2$.

The a priori probability distribution of speckle intensities
implies that the mean number $M$ of speckles (local maxima)  in
volume $V$ with intensities above value $I$ is given by (see Eq.
(21) of Ref. \cite{Garnier1999})
\begin{equation}\label{Idistr}
M(I)=\frac{\pi^{3/2}\sqrt{5}V}{27F^4\lambda_0^3\pi}\left
[\left(\frac{I}{\langle I \rangle}\right
)^{3/2}-\frac{3}{10}\left(\frac{I}{\langle I \rangle}\right
)^{1/2} \right]\exp\left(-\frac{I}{\langle I \rangle}\right ),
\end{equation}
where $\langle I\rangle = P_{beam}/S$ is the average beam
intensity, $S$ is the beam cross section. Then $M(I_c)$ is the
number of collapses per volume $V$ and $P_{scattered}=P_cM(I_c)$
is the optical power scattered out of the main beam  due to
self-focusing. Therefore, rate of scattering is given by
\begin{equation}\label{Pscattered}
P^{-1}_{beam}dP_{scattered}/dz=\frac{P_cM(I_c)}{\langle I \rangle
V}.
\end{equation}
 For NIF parameters,
Eqs. $(\ref{Idistr})$ and $(\ref{Pscattered})$ give
$P^{-1}_{beam}dP_{scattered}/dz=0.0002/\mbox{cm}$ for  $\langle
I\rangle =10^{15}\mbox{W}/\mbox{cm}^2$ and
$P^{-1}_{beam}dP_{scattered}/dz=0.8/\mbox{cm}$ for $\langle
I\rangle =2\times 10^{15}\mbox{W}/\mbox{cm}^2$. If Max's lower
value of $P_c$ were used, order unity of the total beam power
would have been predicted to scatter over a typical NIF plasma
length of $1$cm, even at the lower intensity since
$dP_{scattered}/dz$  is exponentially sensitive (see Eq.
$(\ref{Idistr})$)   to the parameter $\alpha$, with
$dP_{scattered}/dz\propto\exp(-\alpha)$, and $\alpha=I_c/\langle
I\rangle$. For NIF parameters, $\alpha\gg 1$.

\section{Time-dependent self-focusing}

Clearly beam spray due to speckle self-focusing could be a
problem at the higher intensity.  This is alleviated by temporal
beam smoothing techniques
\cite{LehmbergObenschain1983,Skupsky1989} which induce finite
speckle coherence time, $T_c$: the intensity distribution of light
intensity inside plasma is given by a speckle field at each
moment of time as in Figure 1a but location of hot spots changes
in a random manner with a typical time  $T_c$. Such techniques are
used in contemporary experiments \cite{NiemannPRL2005} and in
future experiments at NIF.

Inertia weakens the plasma density response: if $T_c$  is less
than the duration of a particular self-focusing event, $\approx F
\lambda_0 /c_s\sqrt{P/P_c}$, (this estimate is accurate for $P/P_c
\gtrsim 2.5$, see Ref. \cite{RoseDuBoisPhysFluid1993}) then this
self-focusing event will be suppressed. This suppression effect
is significant if $T_c\lesssim F \lambda_0/c_s$, i.e., $T_c$ must
be smaller than the time it takes for a sound wave to cross a
speckle width ($\sim 4$ps for NIF parameters). Here $c_s$ is the
ion-acoustic wave speed. (This is in contrast to the case of
almost instantaneous response of optical Kerr nonlinearity which
is typical for solids \cite{Boyd2002}). As $T_c$ decreases, a
smaller fraction of the beam power participates in collapse
events, controlled by the parameter $\alpha(l_c/c_sT_c)^2$,
instead of $\alpha$, for time independent self-focusing. This has
led to the common assumption \cite{Lindl2004} that if the total
power participating in independent collapse events is made
arbitrarily small by reducing  $T_c$, then beam spray could be
reduced to any desired level.

However, we have found \cite{LushnikovRosePRL2004}  that even for
very small  $T_c$, self-focusing can lead to strong beam spray.
Now, self-focusing results from a collective instability, CFSBS,
which couples the beam to ion acoustic waves that propagate
transversely to the direction of laser beam propagation. As $l_c$
increases, the well-known dispersion relation of forward
stimulated Brillouin scattering \cite{SchmittAfeyan1998} is
recovered for coherent laser beam. We predict that this
instability is not a sensitive function $T_c$  for $c_sT_c\lesssim F
\lambda_0$. Recent experiments at the Omega laser facility
\cite{NiemannPRL2005} are in excellent agreement with that
prediction: It was found that reducing  $T_c$ from 3.4ps (for
which  $c_s T_c\approx F \lambda_0$) to 1.7ps did not cause a
further reduction of beam spray at $\langle I\rangle=5\times
10^{14}\mbox{W}/\mbox{cm}^2$. Note that dominant seed for CFSBS
is not thermal but time-dependent plasma density fluctuations
caused by fluctuating speckles.

\section{Thermal self-focusing}

Quantitative comparison with this data requires extension of our
earlier  work \cite{LushnikovRosePRL2004} to allow transport of
fluctuations, $\delta T_e$, in electron temperature. In that case
the second mechanism of self-focusing - thermal self-focusing
comes into play. Propagation of  laser beam is described by
 paraxial equation for the electric field spatiotemporal
envelope, $E$,
\begin{equation}\label{Eeq1}
  \Big (i\frac{\partial}{\partial
  z}+\frac{1}{2k_0}\nabla^2-\frac{k_0}{2}\frac{n_e}{n_c}\rho
  \Big )E=0, \ \nabla=(\frac{\partial}{\partial x},\frac{\partial}{ \partial
  y}),
\end{equation}
 which is coupled to linearized hydrodynamic equation for the
relative density fluctuation, $\rho=\delta n_e/n_e$, as it
propagates acoustically with acoustic speed $c_s$:
\begin{equation}\label{neq1}
\Big (\frac{\partial^2}{\partial
  t^2}+2\tilde\nu\frac{\partial}{\partial t}-c_s^2\nabla^2  \Big )\ln (1+\rho)=c_s^2 \nabla^2 \Big (I+\frac{\delta
  T_e}{T_e}\Big ),
\end{equation}
where $\delta T_e$ is the fluctuation of electron temperature,
$k_0=2\pi/\lambda_0$, $I=|E|^2$ is the light intensity, $\tilde
\nu$ is an integral operator whose Fourier transform in  $x$ and
$y$ is $\nu_{ia} k c_s$, where  $\nu_{ia}$ is the ion acoustic
wave amplitude damping rate normalized to the ion acoustic
frequency. $x$ and $y$  are transverse directions to beam
propagation direction $z$. $E$ is in thermal units defined so
that in equilibrium, with uniform $E$, the standard $\rho=\exp
(-I)-1$ is recovered. $n_c=m_e\omega_0^2/4\pi e^2$ is the
critical electron density, $m_e$ is the electron mass and $e$ is
the electron charge.
 The relative electron
temperature fluctuation, $\delta T_e/T_e$ is responsible for
thermal self-focusing and was omitted in our previous work
\cite{LushnikovRosePRL2004}.

We make the ansatz that the Fourier transform of electron
temperature fluctuation, $\delta T_e(k)/T_e$ satisfies,
\begin{equation}\label{deltateq1}
\Big (\tau_{ib}\frac{\partial}{\partial
  t}+1  \Big )\frac{\delta T_e(k)}{T_e}=g(k\lambda_e)I(k),
\end{equation}
which is a reduced version of Epperlein's model
\cite{EpperleinShortPhysPlasm1994}.  Here the right-hand-side
(r.h.s.) determines plasma heating by the inverse bremsstrahlung,
$I(k)$ is the Fourier transform of  $I$, so that intensity
fluctuations are a source of $\delta T_e$ \cite{Kruer1990}.  The
inverse bremsstrahlung relaxation time, $\tau_{ib}$, is given by,
\begin{equation}\label{tauib}
\tau_{ib}=\frac{1}{kc_s}\frac{3}{128}\sqrt{\frac{\pi
Z^*\phi}{2}}\frac{\big [1+(30k\lambda_e)^{4/3}\big
]}{k\lambda_e}\frac{c_s}{v_e}.
\end{equation}
Also
\begin{equation}\label{gkl}
g(k\lambda_e)=\frac{\big [1+(30k\lambda_e)^{4/3}\big
]}{96(k\lambda_e)^2}Z^*,
\end{equation}
and $\phi$ is an empirical factor
\cite{EpperleinShortPhysFluid1992}, $\phi=(4.2+Z^*)/(0.24+Z^*)$,
$Z^*=\sum\limits_in_iZ_i^2/\sum\limits_in_iZ_i$ is the effective
plasma ionization number, $n_i$ and $Z_i$ are the number density
and the ionization number (number of ionized electrons per atom)
of $i$-th ion species of plasma, respectively. $\lambda_e$ is
related to the standard $e-i$ mean free path, $\lambda_{ei}$,  by
$\lambda_e=(\lambda_{ei}/3)(2Z^*/\pi\phi)^{1/2}$.  The basic ion
acoustic wave parameters, $\nu_{ia}$ and $c_s$, are regarded as
given by kinetic theory
\cite{BrantovPRL2004,BergerValeoBrunnerPhysPlasm2005,BergerValeoPhysPlasm2005}
which, e.g., takes into account the effect of compressional
heating on sound wave propagation. For comparison with experiment
in this paper, however, collisionless theory is used for
evaluation of acoustic wave parameters.

Eq. $(\ref{deltateq1})$ implies that thermal conductivity is
determined by
\begin{equation}\label{kappa1}
\kappa=\frac{3}{2}\frac{n_e}{\tau_{ib}k^2}=\frac{k_{SH}}{1+(30k\lambda_e)^{4/3}},
\end{equation}
where $k_{SH}$ is the classical Spitzer-Harm
\cite{SpitzerHarm1953} thermal conductivity coefficient in
plasma. Since  $l_c$ is {\it not} small compared to the electron
ion mean free path, $\lambda_{ei}$, thermal transport becomes
nonlocal, and $k_{SH}$   is effectively reduced, as given by Eq.
$(\ref{kappa1})$, when applied to a fluctuation at speckle
wavenumbers, $k=O(1/l_c).$ This reduction of $\kappa_{SH}$ is
substantial for experiment of Ref. \cite{NiemannPRL2005},
implying much larger $\delta T_e$ than classical transport
\cite{EpperleinShortPhysPlasm1994}. Importance of the thermal
contribution to self-focusing at the speckle scale was first
realized by Epperlein
\cite{EpperleinPRL1990,EpperleinShortPhysPlasm1994}, on the basis
of Fokker-Planck simulations, and later analytically derived
\cite{MaximovSilin1993} and verified experimentally
\cite{MontgomeryPRL2000}. It was recently realized
\cite{BrantovPRL2004,BergerValeoBrunnerPhysPlasm2005} that
Epperlein's result
\cite{EpperleinPRL1990,EpperleinShortPhysPlasm1994} is correct
provided the acoustic frequency $c_s/l_c$  is smaller than the
electron-ion collision frequency $v_e/\lambda_{ei}$.

To solve Eqs. $(\ref{Eeq1}),(\ref{neq1})$ and $(\ref{deltateq1})$ we
need to determine boundary conditions on $E$. We assume, absent
plasma, that in the optic far field the Fourier spectrum of $E$
is top-hat with square shape:
\begin{eqnarray}\label{tophatsquare}
|\hat E({\bf k})|=const \ \mbox{for} \ |k_x|<k_m \ \mbox{and} \
|k_y|<k_m; \ |\hat E({\bf k})|=0, \ \mbox{otherwise},
\end{eqnarray}
where $k_m=l_c^{-1}$. Thus our boundary conditions
correspond to square top hat.The superposition of all these Fourier
modes propagating in uniform density plasma we refer to as $E_0$,
the solution of Eq. $(\ref{Eeq1})$ with $\rho=0.$ We assume
temporal beam smoothing which means that Fourier modes $\hat
E({\bf k})$ with different ${\bf k}$ are uncorrelated and the
modes with the same $ \bf k$ are correlated with short
correlation time $T_c< l_c/c_s$.

For NIF designs, $Z^*$ is highly variable depending on details of
plasma composition.  Laser beam may pass through, {\it e.g.},
$He$, $Be$, $CH$, $SiO_2$ and $Au$ plasma, allowing a wide range
of $Z^*$. When $Z^*$ is small, thermal effects are small, and our
previous ponderomotive theory \cite{LushnikovRosePRL2004} applies.
In this case, the linear stage of the collective instability
depends only on one parameter - dimensionless intensity
\cite{LushnikovRosePRL2004},
\begin{eqnarray}\label{Ipondermotive}
\tilde I_{0}=\frac{4F^2}{\nu_{ia}}\frac{n_e}{n_c}
I_0\propto\frac{1}{\alpha\nu_{ia}}.
\end{eqnarray}
 $I_0$ is the spatial average of $|E|^2$.  Note that the standard figure of merit for
self-focusing, $1/\alpha$, is smaller by the factor $\nu_{ia}$
(see Ref. \cite{Lindl2004}).

\section{ collective forward stimulated
Brillouin scatter and transition to beam spray regime}

For small $T_c$, one might expect $\rho\simeq 0$ and that the
laser beam would propagate with $E=E_0$. However, linearization of Eqs.
$(\ref{Eeq1}),(\ref{neq1})$ and $(\ref{deltateq1})$ about this state shows that this propagation
 is unstable.  Following ideas of Ref.
\cite{LushnikovRosePRL2004} and setting $\rho=\delta\rho
e^{\lambda z}\exp\big[i({\bf k}\cdot {\bf x}-\omega t)\big ]$,
$E=E_0+\delta E e^{\lambda z}\exp\big[i({\bf k}\cdot {\bf
x}-\omega t)\big ]$, we obtain the following   dispersion
relation, at acoustic resonance $\omega=kc_s$, assuming  ${\bf k}$ parallel to
either the $x$ or $y$ directions:
\begin{equation}\label{disp1}
2i\nu_{ia}=\Big [1-\frac{g(k\lambda_e)}{1-ikc_s\tau_{ib}}\Big
]\frac{\delta I}{\delta \rho},
\end{equation}
where the plasma density response function $\delta I/\delta
\rho$ is given by
\begin{equation}\label{densresponse}
\frac{\delta I}{\delta
\rho}=\frac{n_e}{n_c}\frac{k_0^2I_0}{4kk_m}\ln
\frac{k^2(-2k_m+k)^2+4k_0^2\lambda^2}{k^2(2k_m+k)^2+4k_0^2\lambda^2}.
\end{equation}
In general case of arbitrary direction of ${\bf k}$ the
dispersion relation is much more bulky and not given here because
it gives essentially the same result.

 Note that for $k_m\to 0 \ (F^2\gg
1
)$ Eq. $(\ref{densresponse})$ reduces to
\begin{equation}\label{densresponsecoherent}
\frac{\delta I}{\delta
\rho}=-\frac{n_e}{n_c}\frac{2k^2k_0^2I_0}{4k_0^2\lambda^2+k^4}
\end{equation}
which means that Eq. $(\ref{disp1})$, absent thermal effects (i.e. for $\delta T_e=0$ in Eq.
$((\ref{neq1})$),
reduces to the paraxial limit of the standard FSBS dispersion
relation \cite{SchmittAfeyan1998}.

Absent thermal effects we regain the pondermotive case considered
in Ref.  \cite{LushnikovRosePRL2004} except that
in this Article square top hat boundary conditions
$(\ref{tophatsquare})$ are used compared with circular top hat boundary
conditions used in \cite{LushnikovRosePRL2004}. We find however
that both circular and square top hat boundary conditions give
similar results.

Positive value of $Re(\lambda)$ corresponds to convective
instability so that the fluctuations of beam intensity grow as
$\exp\big[ Re(\lambda) z\big ]$  with distance.  When $\lambda$
is non-dimensionalized, $\tilde \lambda=l_c^2k_0\lambda$, it only
depends on
\begin{equation}\label{Itilde}
{\tilde I}\equiv\Big [1+\frac{g(k\lambda_e)}{1-ikc_{s}\tau_{ib}}\Big
] \tilde I_{0}.
\end{equation}
 Here $k$ is determined from the condition that
$Re(\lambda)$ is  maximum and $\tilde I_0$ is given by
$(\ref{Ipondermotive})$. According to our theory of CFSBS,
$\lambda^{-1}$, should be compared with the basic correlation
length in $z$ direction, known as the speckle length,
$l_{speckle}\approx 7F^2\lambda_0$. The value $\tilde
\lambda=0.1$, at which $\lambda\approx l_{speckle}$, marks regime
transition. In the first, weak regime, with $\tilde \lambda\ll
0.1$, there is little gain over a speckle length. It follows that
only small changes in correlations develop over a speckle length,
in particular, there is little change in $\triangle \theta$.
Changes over different speckles are uncorrelated, leading to a
quasi-equilibrium (see Figure 3 of Ref.
\cite{LushnikovRosePRL2004}). As $\tilde \lambda$  crosses the
value $0.1$ (corresponding to $\tilde I\approx 2$  in
ponderomotive case), a second, non-equilibrium regime, is
entered, and beam properties change rapidly with $z$. In
particular, $\triangle \theta$  changes rapidly, {\it i.e.},
there is beam spray. This is shown in Figure 2, where normalized
beam spray rate is shown.
\begin{figure}
\begin{center}
\includegraphics[width = 5.4 in]{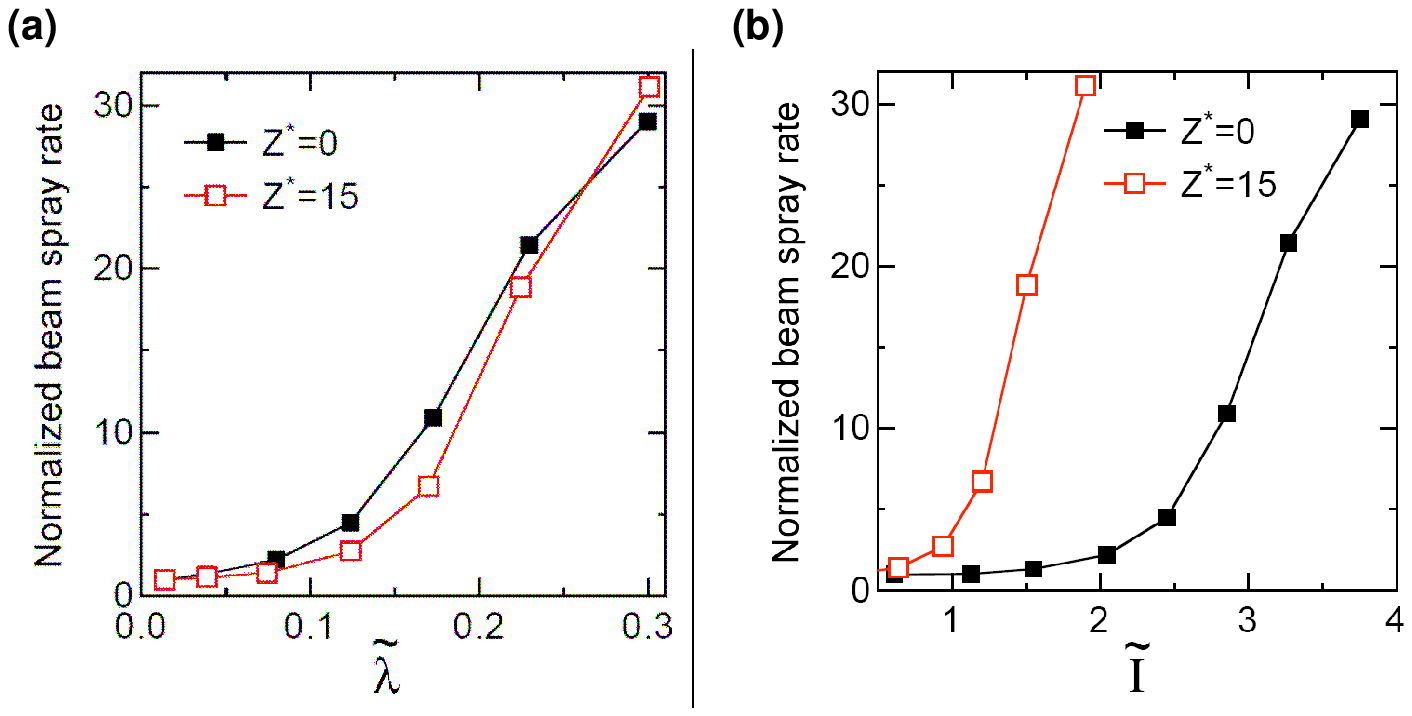}
\caption{Figure 2. Dependence of beam spray rate obtained from
simulations as a function of  (a) dimensionless growth rate
$\tilde \lambda$ and (b) dimensionless intensity $\tilde I$.  Red
curves correspond to ponderomotive self-focusing ($Z^*=0$) and
black curves ($Z^*=15$) correspond to case there both
ponderomotive and thermal self-focusing are essential. Both red
and black curves collapse to the single curve in (a) which
indicated that $\tilde \lambda$  is a much better parameter for
onset of beam spray, compared with  $\tilde I$.} \label{fig:fig2}
\end{center}
\end{figure}
 Note that absent instability one expects
beam spray rate  $d\langle \theta^2 \rangle/dz\sim I_0^2.$ So in
Figure 2 we normalized beam spray rate to $I_0^2$ (see Ref.
\cite{LushnikovRosePRL2004} for more discussion). Compared with
Figure 6 of Ref. \cite{LushnikovRosePRL2004}, there has been an
important change of independent variable, from $\tilde I$ to
$\tilde \lambda$, which allows a unified presentation of both
ponderomotive and thermal cases.

Thus analysis of $\tilde \lambda$  results in  second and main
conclusion of our CFSBS theory: prediction of the onset of beam
spray, and hence a prediction of fundamental limit on power
propagation. Here we present comparison of this prediction with
\cite{NiemannPRL2005}, the first experimental measurement of beam
spray onset (see Figure 3).
\begin{figure}
\begin{center}
\includegraphics[width = 4.4 in]{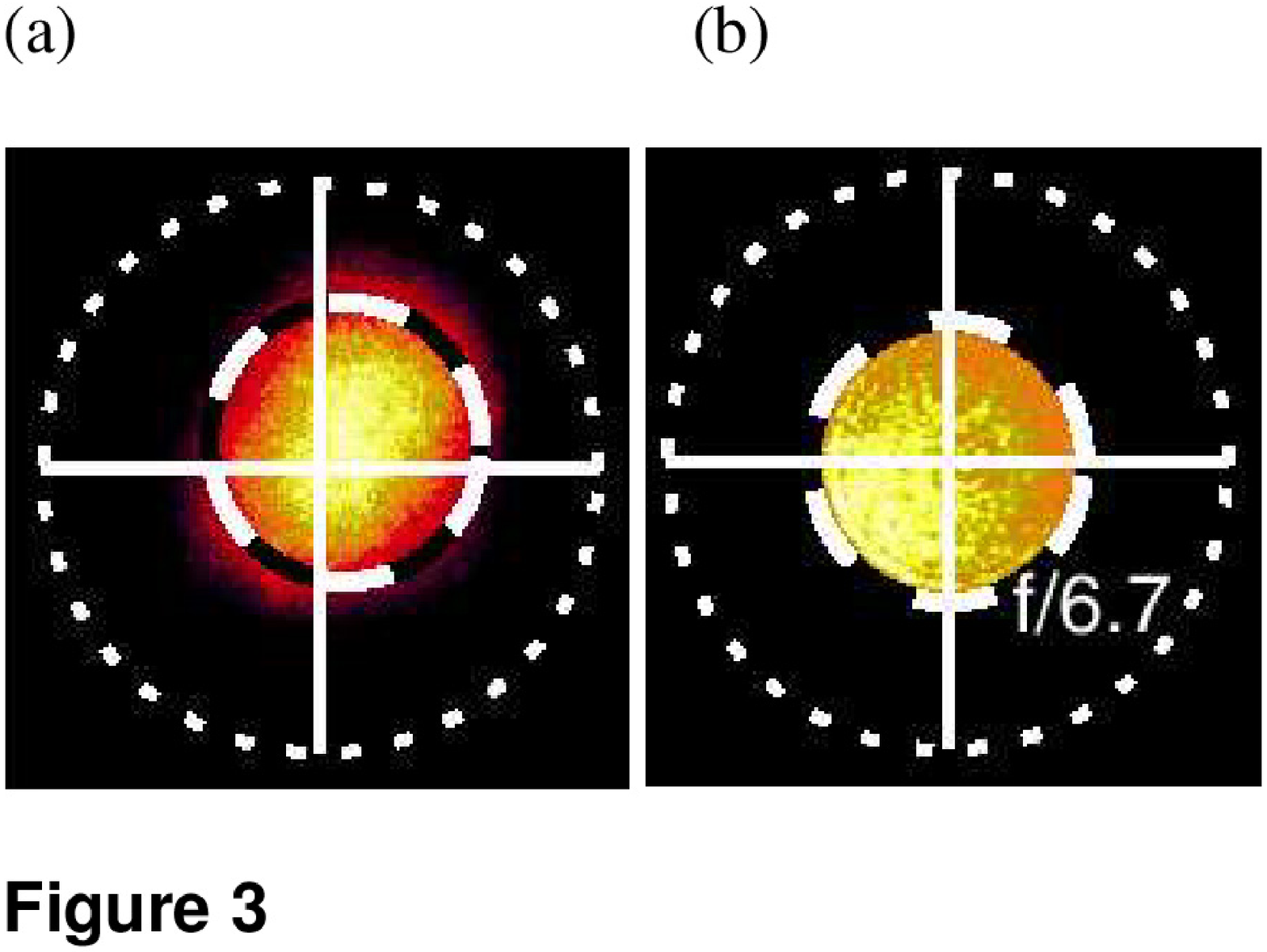}
\caption{Figure 3. Experimental images of cross section of  time
averaged laser beam intensity after propagation through plasma.
(a) Onset of beam spray regime at $5\times
10^{14}\mbox{W}/\mbox{cm}^2$. (b) Negligible beam spray regime
achieved by lowering intensity. Dashed circles correspond to
$F=6.7$ beam width for propagation in vacuum. Reproduced from
\cite{NiemannPRL2005} with permission.} \label{fig:fig3}
\end{center}
\end{figure}
 From
\cite{NiemannPRL2005,NiemannPrivate,MeezanPrivate} we find that
$0.14<n_e/n_c<0.25. \ T_e\sim 2\mbox{keV}, \ F=6.7,\
\omega_0\approx3.6\times10^{15}\mbox{sec}^{-1}$, and $Z^*=6.4$ at
upper range of densities.
 For a nominal electron density of  $n_e=0.2n_c$, the $0.1$ contour
 (color online) of $\tilde \lambda$  is shown in
  Figure 4a, implying $\tilde I\approx0.65$  at regime transition.
\begin{figure}
\begin{center}
\includegraphics[width = 4.4 in]{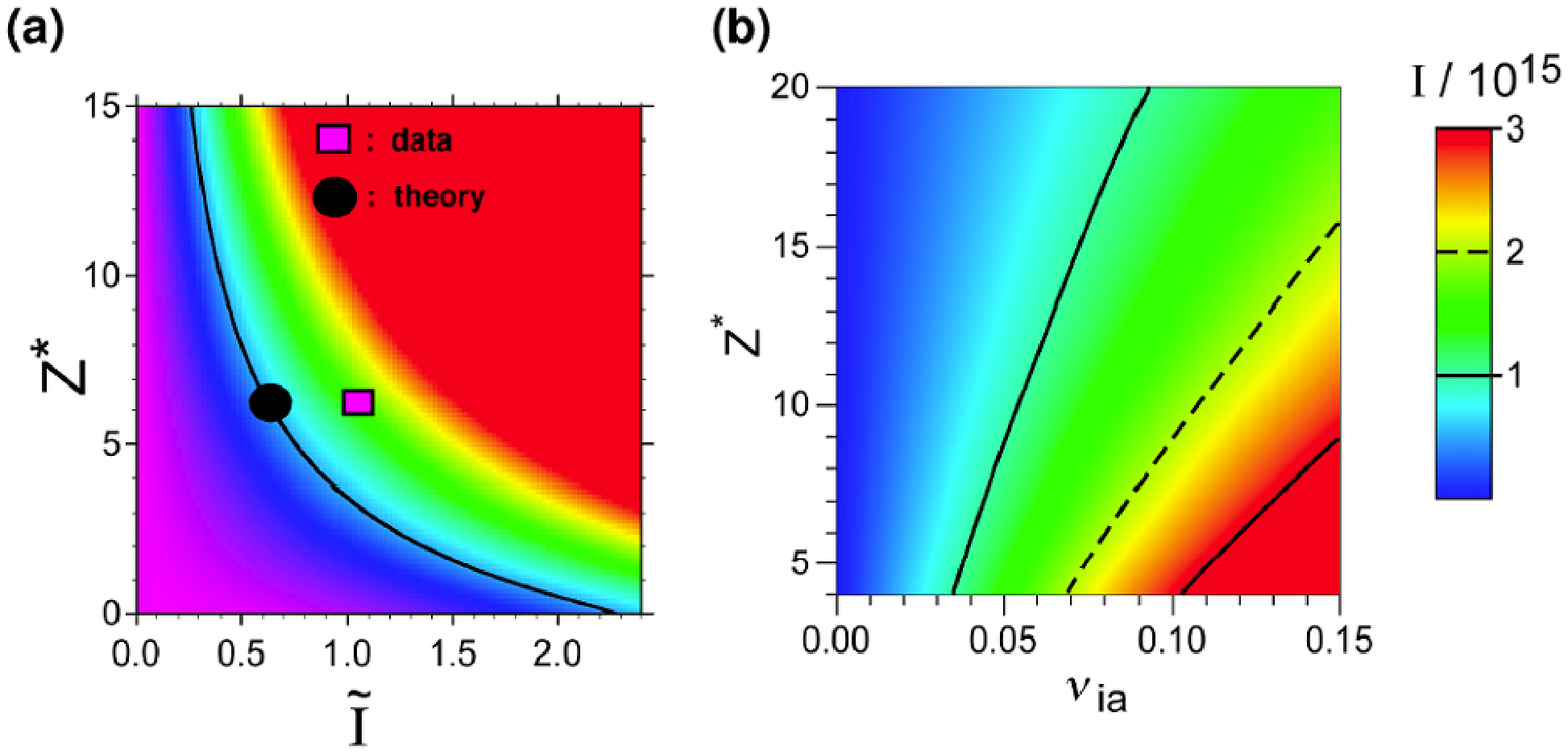}
\caption{Figure 4. (a) Solid curve separates predicted beam spray
regime, $\tilde \lambda>0.1$  (from green to red colours), from
negligible beam spray regime $\tilde \lambda<0.1$, (from blue to
purple colours). Different colours denote values of  $\tilde
\lambda$, with red corresponding to the value $0.3$ and above.
Magenta square denotes experimentally measured (Ref.
\cite{NiemannPRL2005}) beam spray onset, assuming $\nu_{ia}  =
0.06$ and black circle is the theoretical prediction for $\nu_{ia}
= 0.06$.  (b) Predicted onset of beam spray regime ({\it i.e.}
for $\tilde \lambda =0.1$) as a function of $Z^*$ and $\nu_{ia}$
for NIF plasma with $T_e\sim 5\mbox{keV}, \ F=8, \ n_e/n_c=0.1, \
\omega_0\approx 5.4\times 10^{15}\mbox{sec}^{-1}$. Colours show
laser intensity, in units of $10^{15} \mbox{W}/\mbox{cm}^2$.
Intensity is at maximum for small $Z^*$ and large $\nu_{ia}$.   We
assume $Z^*>4$ to make sure that condition
$c_s/l_c<v_e/\lambda_{ei}$ is true.} \label{fig:fig4}
\end{center}
\end{figure}
  was observed \cite{NiemannPRL2005}, corresponds to  $\tilde I\approx1.05$, with Landau damping $\nu_{ia}=0.06$  for the plasma composition at
  this density. The major uncertainty in comparing this data with theory is due to significant time dependence of $T_e/T_i$ during experiments as well as plasma
  density inhomogeneity, {\it e.g.}, if $n_e=0.14n_c$  (which corresponds to plasma density plateau in
  Figure 3 of Ref. \cite{NiemannPRL2005}) with other parameters the same, then theory predicts $\tilde I\approx0.73$  and experiments give  $\tilde I\approx 0.82$.
In contrast, prediction based on speckle collapses, gives that
even at the maximum density of  $n_e/n_c=0.25, \
P^{-1}_{beam}dP_{scattered}/dz=0.23\mbox{cm}^{-1}$, the scattered
power fraction, $P_{scattered}/P_{beam}$, is only $0.5\%$
 after $200\mu m$ of propagation through the high-density region of the plasma.
  This is much less then the observed \cite{NiemannPRL2005} $10\%$.
  Therefore, beam spray due to CFSBS is consistent with the data while beam spray due to speckle collapse is not.

\section{Implication for backscattering}

Recent experiments at the Atomic Weapon Establishment in the UK
have demonstrated reduction of both stimulated Brillouin and
Raman backscatter
 \cite{Suter2004} by the addition of small amounts of high ionization
state dopants to a low ionization state plasma, e.g., a $1\%$
dopant reduced backscatter by more than an order of magnitude.
Combination of these experimental facts  with our prediction that
dopant may cause transition to beam spray regime suggests that one
should expect anti-correlation between beam spray and
backscatter. If this anti-correlation is confirmed experimentally
then we propose the following mechanism:  beam spray decreases
speckle length (correlation length) with beam propagation and
backscatter is suppressed by reduction of laser beam correlation
length. The latter has been established through simulation
\cite{RoseDuBoisPRL1994}, experiment \cite{FernandezPRE1996} and
one dimensional analytic theory \cite{MounaixPhysPlasm1995}. In
other words, control of backscatter is achieved indirectly through
control of CFSBS. We are unaware of any other explanation of this
backscatter reduction by the addition of small amounts of high Z
dopant.

Clearly, to maintain control of forward beam propagation, beam spray must not be strong.
If plasma paramters are conducive to
backscatter as in the Atomic Weapon Establishment experiment \cite{Suter2004}, then by
altering the plasma state so as to be above, but close to, the beam spray regime transition, allowing
moderate beam spray might lead to optimum control of beam propagation and backscatter.
This suggests operating above but, {\it e.g.}, close to the solid curve of figure 4a which
marks the transition regime of CFSBS.

\section{Conclusion}

In conclusion, transition to
 the beam spray regime was recognized as a collective phenomenon. Our theory is in excellent agreement with
experiment: the transition laser intensity   and its insensitivity
 to changes in correlation time were predicted. We found that the growth rate of CFSBS depends on
four dimensionless parameters: the scaled laser intensity  $\tilde
I_0$ (see Eq. $(\ref{Ipondermotive})$), scaled electron-ion mean
free path $\lambda_{ei}/F\lambda_0$, effective ionization number
$Z^*$, and  $c_s/v_e$. The first three of these can be
manipulated experimentally. So our theory permits predictions
 for beam control at NIF that may be implemented since thermal self-focusing can be
 manipulated experimentally through control of CFSBS in two ways.  First, by changing $Z^*$  through change of plasma composition.
 For example, addition of $1\%$ of Xenon (high $Z$ dopant) to low $Z$ plasma
  ($50\%$ of $He$ and $50\%$ of $H$) would increase $Z^*$  from $1.7$ to $15.5$ without significant change in $\nu_{ia}$.
Second, beam control can be implemented by adding low $Z$ dopant to a high $Z$ plasma, e.g., adding $He$ to $Si0_2$, in order to increase
$\nu_{ia}$ at almost constant $Z^*$.  Figure 4b shows dependence
of laser intensity (indicated by colors) at predicted onset of
beam spray regime on $Z^*$ and $\nu_{ia}$  for NIF parameters. It
is seen that maximal allowable intensity occurs for small $Z^*$
and large $\nu_{ia}$. We propose Figure 4b as direct guide for
choice of NIF designs to attain maximum power of laser beam,
which may propagate without significant beam  spray.

Observation of anti-correlation between beam spray and
backscatter, through the addtion of small amounts of high Z
dopant, would mean addtional confirmation of our theory. We
predict that  control of backscatter is achieved indirectly
through control of CFSBS, e.g. by changing plasma ionization
state and/or acoustic damping \cite{FernandezPhysPlasm1997}.

{\bf Acknowledgements}  We thank R.L. Berger for attracting our
attention to Refs.
\cite{BrantovPRL2004,BergerValeoBrunnerPhysPlasm2005}
 and pointing
out that seed for CFSBS provided by the fluctuating speckles is
much larger than thermal. We thank W. Rozmus for pointing out the
limitation of Epperlein's model
\cite{EpperleinPRL1990,EpperleinShortPhysPlasm1994}
 to
$c_s/l_c<v_e/\lambda_{ei}$. Support was provided by the
Department of Energy, under contract W-7405-ENG-36.

{\bf Correspondence} and requests for materials should be
addressed to H.R. (har@lanl.gov).



\begin{thebibliography}{}




\bibitem{McCrory1988} McCrory R L et. al. 1988 Nature {\bf 335} 225

\bibitem{Still2000} Still C H et al. 2000 Physics of Plasmas {\bf 7} 2023

\bibitem{Miller2004} Miller G H Moses E I \& Wuest C R 2004 Nucl. Fusion {\bf 44} S228

\bibitem{Lindl2004} Lindl J D et. al. 2004 Phys. Plasmas {\bf 11}, 339

\bibitem{Boyd2002} Boyd RW 2002 Nonlinear Optics (Academic Press, San Diego)

\bibitem{Sulem1999} Sulem C and Sulem P L 1999 Nonlinear Schroedinger Equations:
Self-Focusing and Wave Collapse (Springer)

\bibitem{LushnikovRosePRL2004} Lushnikov P M and Rose H A 2004 Phys. Rev. Lett. {\bf 92}  255003

\bibitem{NiemannPRL2005} Niemann C et. al., 2005 Phys. Rev. Lett. {\bf 94} 085005

\bibitem{Kruer1990} Kruer W L 1990 The physics of laser plasma interactions.
Addison-Wesley, New York)

\bibitem{ZakharovShabat1971} Zakharov V E and Shabat A B 1971 Zh. Eksp. Teor. Fiz.,  {\bf
61} 118  [1972 Sov. Phys. JETP, {\bf 34}, 62]

\bibitem{ChiaoGarmireTownes1964} Chiao R Y, Garmire E and Townes C H 1964 Phys. Rev. Lett.
{\bf 13} 479

\bibitem{TalanovLETPLett1965} Talanov V I 1965 JETP Letters {\bf 2} 138

\bibitem{RoseDuBoisPRL1994} Rose H A and DuBois D F 1994 Phys. Rev. Lett. {\bf 72} 2883-2886

\bibitem{Max1976} Max C E 1976 Phys. Fluids {\bf 19} 74

\bibitem{RoseDuBoisPhysFluid1993} Rose H A and DuBois D 1993 Phys. Fluids B {\bf 5} 3337

\bibitem{KatoMima1982} Kato Y and Mima K 1982 Appl. Phys. B {\bf 29} 186

\bibitem{Garnier1999} Garnier J 1999 Phys. Plasmas {\bf 6} 1601 Eq. (21).

\bibitem{LehmbergObenschain1983} Lehmberg R H and Obenschain S P 1983 Opt. Commun. {\bf 46} 27

\bibitem{Skupsky1989} Skupsky S et. al., 1989 J. Appl. Phys. {\bf 66} 3456

\bibitem{SchmittAfeyan1998} Schmitt A J and Afeyan B B 1998 Phys. Plasmas {\bf 5} 503

\bibitem{SpitzerHarm1953} Spitzer L Jr and Harm R 1953 Phys. Rev. {\bf 89} 977

\bibitem{EpperleinPRL1990} Epperlein E M 1990 Phys. Rev. Lett. {\bf 65} 2145

\bibitem{EpperleinShortPhysFluid1992}Epperlein E M and Short R W 1992 Phys. Fluids {\bf B4}, 2211

\bibitem{MaximovSilin1993} Maximov AV and Silin V P 1993 Zh. Eksp. Teor. Fiz., {\bf 103} 73
[1993 Sov. Phys. JETP {\bf 76} 39]

\bibitem{EpperleinShortPhysPlasm1994} Epperlein E M and Short R W 1994 Phys. Plasmas. {\bf 1} 3003

\bibitem{MontgomeryPRL2000} Montgomery D S,  Johnson R P, Rose H A, Cobble J A and
Fernandez J C 2000 Phys. Rev. Lett. {\bf 84} 678

\bibitem{BrantovPRL2004} Brantov A V et. al., 2004 Phys. Rev. Lett. {\bf 93} 125002

\bibitem{BergerValeoBrunnerPhysPlasm2005}
Berger R L,  Valeo E J, and Brunner S  2005 Phys. Plasmas {\bf 12}
062508

\bibitem{BergerValeoPhysPlasm2005}
Berger R L,  and Valeo E J  2005 Phys. Plasmas {\bf 12} 032104

\bibitem{NiemannPrivate} We thank C. Niemann for communicating to us the detailed
plasma composition.

\bibitem{MeezanPrivate}  Determined by simulation, results
provided by N.B. Meezan, private comm. (2005)

\bibitem{Suter2004} Suter L J et. al.  2004 Phys. Plasmas {\bf 11} 2738

\bibitem{FernandezPRE1996}Fernandez J C et. al. 1996 Phys. Rev. E {\bf 53} 2747

\bibitem{MounaixPhysPlasm1995} Mounaix Ph. 1995 Phys. Plasmas {\bf 2} 1804

\bibitem{FernandezPhysPlasm1997}Reduction of Stumulated Raman bacscatter by reducing acoustic damping has been observed in Fernandez J C et. al. 1997 Phys. Plasm. {\bf 4}
1849

\end{thebibliography}
\end{document}